\documentclass[twocolumn,amsmath,amssymb]{revtex4}
\usepackage{amsmath}
\usepackage{amssymb}
\usepackage{graphicx}

\usepackage{color}
\usepackage{ulem}

\begin{document}
\title{Spin liquid and infinitesimal-disorder-driven cluster spin glass 
in the kagome lattice}

\author{M. Schmidt$^1$}
\email[]{mateusing85@gmail.com}
\author{F. M. Zimmer$^1$}
\email[]{fabiozimmer@gmail.com}
\author{ S. G. Magalhaes$^2$} \email[]{sgmagal@gmail.com}

\affiliation{$^1$ Departamento de F\'isica, Universidade Federal de Santa Maria, 97105-900, Santa
Maria, RS, Brazil \\ 
$^2$ Instituto de F\'isica, Universidade Federal do Rio Grande do Sul,  91501-970, Porto Alegre, RS, Brazil}

\begin{abstract}
The interplay between geometric frustration (GF) and bond disorder is studied  in the Ising kagome lattice within a cluster approach. 
The model considers antiferromagnetic (AF) short-range couplings and long-range intercluster disordered interactions. 
The replica formalism is used to obtain an effective single cluster model from where the thermodynamics is analyzed by exact diagonalization. 
We found that the presence of GF can introduce cluster freezing at very low levels of disorder. The system exhibits an entropy plateau followed by a large entropy drop close to the freezing temperature.
In this scenario, a spin-liquid (SL)
behavior prevents conventional long-range order, but an infinitesimal disorder picks out uncompensated cluster states from the multi degenerate SL regime, potentializing the intercluster disordered coupling and bringing the cluster spin-glass state. 
To summarize, our results suggest that the SL state combined with low levels of disorder can activate small clusters, providing hypersensitivity to the freezing process in geometrically frustrated materials and playing a key role in the glassy stabilization.
We propose that this physical mechanism could be present in several geometrically frustrated materials. In particular, we discuss our results in connection to the recent experimental investigations of the Ising kagome compound Co$_3$Mg(OH)$_6$Cl$_2$.
\end{abstract}
\maketitle 

\section{Introduction}

Magnetic systems with competing interactions present a richness of physical properties that can emerge from a conflicting situation, called frustration \cite{Ramirez_GF_1994, Balents_SL}. 
For instance, when disorder brings frustration, a spin-glass (SG) state can be found with the magnetic moments frozen in random directions \cite{Binder86, Scripta_Nordblad_SG_Review, Mydosh_Review_2015}. The lattice geometric features can also carry frustration \cite{Lacroix_book}.
In particular, 
magnets with geometrical frustration (GF) have been a central topic in condensed matter physics due to the possible realization of exotic states of matter, such as classical or quantum spin liquids (SL) \cite{Balents_SL,Mendels2016455}.
In the last years, experimental efforts in the pursuit of SL materials have revealed a wide range of exciting problems.
A particularly interesting issue occurs when the SG behavior is found in geometrically frustrated materials at very low levels of disorder,  or even, apparently, with disorder-free.
In this case,
there is a number of open questions regarding the interplay  among GF, disorder and glassiness. 
For instance, 
can the arising of a cluster spin-glass (CSG) state be related with 
the SL occurrence? 

Recently, low-temperature short-range correlations have been reported in a plenty of geometrically frustrated magnets \cite{PhysRevB.65.220406,PhysRevB.79.224407, JPCM_Maji_2011,PhysRevB.83.024405,zheng2012,JPSJ_2013_Hanashima,sampathCa3Co2O6,JPCM_Chakrabarty,JPCM_CHakrabarty_2,PhysRevLett.113.117201,PhysRevB.91.054423,Chandragiri201626,PhysRevB.93.014433,sampathMnRSbO}.
In several of these materials, as ZnCr$_2$O$_4$ \cite{PhysRevB.65.220406},
CaBaFe$_{4}$O$_7$ \cite{PhysRevB.79.224407},  Nd$_5$Ge$_3$ \cite{JPCM_Maji_2011}, Y$_{0.5}$Ca$_{0.5}$BaCo$_{4}$O$_7$ \cite{PhysRevB.83.024405}, CoAl$_{2}$O$_{4}$ \cite{JPSJ_2013_Hanashima}, Ca$_{3}$Co$_{2}$O$_6$ \cite{sampathCa3Co2O6} and FeAl$_{2}$O$_{4}$ \cite{PhysRevB.91.054423}, some degree of glassiness is observed.
However, the microscopic signatures are often different from the canonical spin glasses. In particular, a CSG state, in which clusters of spins behave as magnetic unities, is the ground-state of various geometrically frustrated systems \cite{PhysRevB.79.224407, JPCM_Maji_2011, JPCM_Chakrabarty, JPCM_CHakrabarty_2}. 
For this class of systems, it has been proposed that the presence of clusters can provide hypersensitivity to disorder  \cite{PhysRevB.65.220406}. In this way, the composite spin degrees of freedom related to the presence of clusters could play a significant role in the physics of geometrically frustrated magnets.

The kagome lattice structure is one of the most promising candidates for experimental SL \cite{Balents_SL}. Among the many proposed realizations of the kagome lattice, an interesting result is provided by the Ising antiferromagnet Co$_3$Mg(OH)$_6$Cl$_2$ compound. In this material,  signatures of a collective paramagnetic SL state and spin freezing are observed at low temperatures. 
As the temperature decreases, it exhibits a plateau in the entropy curve followed by an entropy drop related to the onset of the SG state \cite{zheng2012}.
However, the source of glassiness remains unclear. The large spin-flipping time suggests that the freezing behavior may not be well described as a conventional SG. Moreover, the neutron diffraction and muon spin rotation/relaxation results could support the presence of small spin clusters. Therefore, the interesting physics reported from this kagome compound still lacks a proper explanation.

From the theoretical side, there are few contributions to account the interplay of glassiness and GF. For instance, it was proposed that a disorder-free SG state can occur in geometrically frustrated systems \cite{PhysRevB.86.024434, free_disorder_Chandra}, accounting for the spin glasses with no measurable disorder. In this framework, the energy barriers associated to the SG behavior could be introduced by GF \cite{free_disorder_Chandra}.
A different perspective is based on the fact that quenched disorder cannot be completely avoided in real materials.
In this way, recent studies on the pyrochlore lattice reported analytical and numerical evidences that a SG ground-state can be induced by very low levels of bond disorder \cite{PhysRevLett.98.157201, PhysRevB.81.014406} or a small amount of randomly distributed nonmagnetic impurities \cite{PhysRevLett.114.247207}.
The freezing temperature ($T_f$) is found to be proportional to the amplitude of disorder strength  \cite{PhysRevB.81.014406} or the dilution of impurities \cite{PhysRevLett.114.247207}, respectively.
In addition, a cluster disordered approach was proposed to the study of the frustrated square lattice \cite{prezimmer2014}. By tuning the ratio between first- and second-neighbor interactions, the authors of Ref. \cite{prezimmer2014} found that a SG state can be observed at any amount of intercluster disorder when GF is present. Nonetheless, it was assumed absence of conventional long-range ordering, by considering only intracluster short-range couplings.

However, novel techniques and mathematical frameworks are still needed to account for the SG state in geometrically frustrated systems. 
In this work, we study the antiferromagnetic Ising kagome system to investigate the onset of a low temperature cluster spin-glass phase, which is suggested to appear in Co$_3$Mg(OH)$_6$Cl$_2$. To accomplish that, we propose a disordered cluster model that considers random gaussian deviations in the AF exchange interactions between clusters. In fact, it allows us to take into account GF and disorder effects within a theoretical framework based on analytical calculations.
 In this approach, the intercluster disorder can introduce a relevant degree of freedom - the cluster magnetic moment - dependent on the AF interactions. In particular, we study this model for the square and kagome lattices, which helps us to compare results with and without GF effects. 
For instance, AF interactions in the square lattice can stabilize the N\'eel state, avoiding the CSG behavior.
On the other hand, GF avoids conventional ordering and can lead to uncompensated clusters, that can be a fundamental ingredient to the onset of the cluster freezing at very low levels of disorder.
In fact, we found that GF prevents N\'eel order in the kagome lattice, driving the SL behavior and allowing the CSG onset at low temperatures.
Even an infinitesimal disorder picks out uncompensated cluster states from the multi degenerate SL regime, potentializing the intercluster disordered coupling and bringing the CSG state.

In order to deal with this problem, we adapt the correlated cluster mean field (CCMF) theory \cite{Yamamoto} to the replica formalism. The replicas are used to evaluate the intercluster disordered interactions by 
a mean-field theory \cite{Soukoulis78-1}. The resulting model is then treated with the CCMF method that considers finite clusters, where the short-range intercluster interactions are replaced by self-consistent mean-fields dependent on the cluster spin configurations \cite{Yamamoto}. 
The CCMF theory takes into account lattice geometric features and catch properly short-range correlations. Furthermore, it provides very accurate results for critical quantities and thermodynamic properties  \cite{Yamamoto,PhysRevE.89.062117, PhysRevE.93.062116}.

This paper is structured as follows. In Sec. \ref{model} we discuss the model and the analytic calculations for the disordered kagome and square systems. Our results are presented in Sec. \ref{res}. In Sec. \ref{conc} we present the conclusion.

\section{Model}\label{model}

We consider the Ising model $H = −\sum_{i,j} J_{ij}\sigma_{i} \sigma_{j}$ with spins $\sigma_i$ on sites $i$ of a regular lattice, in which the exchange interaction is given by  $J_{ij}=J_0+\delta J_{ij}$ 
with the random deviation $\delta J_{ij}$ introducing bond disorder.
Without loss of generality, we write the system Hamiltonian dividing it
into $N_{cl}$ identical clusters of $n_s$ spins each ($N=N_{cl}n_s$): \begin{equation}
 H=  - \sum_{\nu}^{N_{cl}} \sum_{ i,j }^{n_s}J_{i_\nu j_\nu}\sigma_{i_\nu} \sigma_{j_\nu}
 -\sum_{\nu, \lambda}^{N_{cl}}\sum_{i,j}^{n_s}J_{i_\nu j_\lambda} \sigma_{i_\nu} \sigma_{j_\lambda} ,
\label{eq1}\end{equation}
where $\nu$ ($\lambda$) corresponds to the cluster index and $\sigma_{i_\nu}$ represents the Ising spin on the site $i$ at the cluster $\nu$.
This model presents two types of interactions: intracluster and intercluster.
We assume that the deviations in the intracluster couplings (first term of Eq. (\ref{eq1})) can be neglected. 
In this way, the second term of Eq. (\ref{eq1}) retains all the relevant disorder of our approach. Furthermore, the intercluster disorder is evaluated in a mean-field spirit, with all spins of a given cluster under the same disordered coupling: $\delta J_{i_\nu j_\lambda} \approx\delta J_{\nu \lambda}$.  In other words, our approach consider a competition between long-range and short-range interactions, in which we expect that the disorder (long-range) could mimic effects of interactions coming, e.g., from intralayer and interlayer perturbations.

 \begin{figure}
\center{\includegraphics[angle=0,width=0.99\columnwidth]{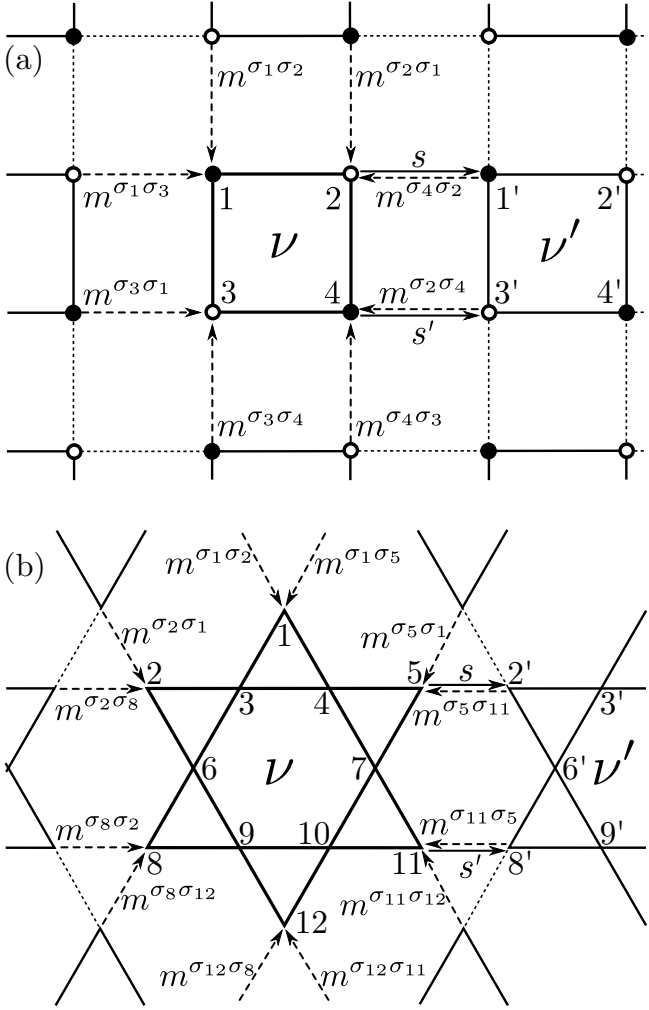}}
\caption{Schematic representation for (a) square and (b) kagome lattices divided into clusters with $n_s=4$ and $n_s=12$, respectively. The mean fields are pointed by dashed arrows that represent the interactions between cluster $\nu$ with its neighbors. The solid arrows indicate  interactions between the clusters $\nu'$ and $\nu$ replaced by $s$ and $s'$ to evaluate the mean fields. For the AF square lattice, it is considered two sublattices represented by solid and open circles.}  
\label{CCMF}  \end{figure}
 
 The disordered cluster model becomes  
\begin{eqnarray}
H &=&  -\sum_{\nu}^{N_{cl}}\sum_{(i,j)}^{n_s}J_0 \sigma_{i_\nu} \sigma_{j_\nu}  
- \sum_{(i_\nu, j_\lambda)} J_{0}\sigma_{i_\nu} \sigma_{j_\lambda}
 \nonumber \\ && - \sum_{(\nu,\lambda)} \delta J_{\nu \lambda}\sigma_{\nu} \sigma_{\lambda},
 \label{eq2}\end{eqnarray}
where $(\cdots)$ represents nearest-neighbors sum and $\sigma_{\nu}=\sum_{i}^{n_s}\sigma_{\nu_i}$ is the total magnetic moment of cluster $\nu$.
The cluster model (Eq. \ref{eq2}) considers uniform antiferromagnetic interactions $J_0$ between nearest-neighbor spins and  disordered couplings only among pairs of spins of neighbor clusters ($\delta J_{\nu\lambda}$). The coupling constants $\delta J_{\nu\lambda}$ follow independent Gaussian distributions with average zero and variance $\bar{J}^2$.

The replica method is used to obtain the free energy average over the quenched random variables: $f=\langle f(\{J_{\nu\lambda}\})\rangle_{J_{\nu\lambda}}=-T/N \lim_{n\rightarrow 0} (\langle Z^n \rangle_{J_{\nu\lambda}}-1)/n$. This procedure consists in evaluating the average of the n-replicated partition function $Z^{n}$, which can be expressed as 
\begin{equation}
 \langle Z^n \rangle_{J_{\nu\lambda}}= \mbox{Tr}\exp(-\beta H_{av})
\end{equation}
with the replicated model
\begin{equation} H_{av}=\sum_{\alpha}
H_{J_0}^{\alpha}-\frac{\bar{J}^2\beta}{2} \sum_{\alpha\gamma}\sum_{(\nu,\lambda)}\sigma^{\alpha}_{\nu} \sigma^{\gamma}_{\nu} \sigma^{\alpha}_{\lambda} \sigma^{\gamma}_{\lambda} ,
\end{equation}
where $H_{J_0}^{\alpha}$ corresponds to the first and second terms of the right side of Eq. (\ref{eq2}) with the replica index $\alpha$.
This problem can be analytically computed in a mean-field approximation  by introducing the variational parameters $q_{\alpha\alpha}=\langle \sigma_\nu^\alpha\sigma_\nu^\alpha\rangle_{H_{av}}$ and $q_{\alpha\gamma}=\langle \sigma_\nu^\alpha\sigma_\nu^\gamma\rangle_{H_{av}}$ ($\alpha\neq \gamma$), where $\langle\cdots\rangle_{H_{av}}$ represents the thermal average over the model $H_{av}$. 
Physically, $q_{\alpha\gamma}$ and $q_{\alpha\alpha}$ correspond to the cluster spin-glass order parameter and the cluster magnetic moment self-interaction, respectively.
This procedure results in the following free energy:
\begin{eqnarray}
f &=& \lim_{n\rightarrow 0} \left[ \frac{\beta J^2}{4n}\left(\sum_{\alpha}{q}_{\alpha\alpha}^2+\sum_{\alpha\gamma}q_{\alpha\gamma}^2\right) \right.
\nonumber \\ && \left.-\frac{\ln \mbox{Tr e}^{-\beta H_{eff}}}{N_{cl}n}\right],
\label{freeenergy}
\end{eqnarray} 
where
\begin{eqnarray}
 H_{eff}&=& - \sum_{\alpha} H^{\alpha}_{J_0} 
 -\frac{\beta J^2}{2}\sum_{\nu}\left[\sum_{\alpha}{q}_{\alpha\alpha}(\sigma_{\nu}^{\alpha})^2
\right. \nonumber  \\  && \left. +\sum_{\alpha\gamma}\frac{q_{\alpha\gamma}}{2}\sigma_{\nu}^{\alpha}\sigma_{\nu}^{\gamma}\right]
  \label{Hav}\end{eqnarray}
with $J=\bar{J}\sqrt{z}$ ($z$ is the number of neighbor clusters) and $q_{\alpha\gamma}$ and $q_{\alpha\alpha}$ are obtained from the extreme condition of free energy.
At this point, there is still a coupling between spins of neighbor clusters at the same replica (first term of Eq. \ref{Hav}). For this intercluster replica coupling, the present work adopts the framework of the CCMF approach, that allows us to decouple the clusters by treating the remaining interactions with good accuracy \cite{Yamamoto}. This procedure results in the following effective single cluster model (see 
\ref{CCMF_dec}):
\begin{eqnarray}
H_{eff}&=& -\sum_{\alpha}\left[\sum_{(i,j)}  J_0\sigma_{i_\nu}^{\alpha}\sigma_{j_\nu}^{\alpha} +
\sum_{i\in \bar{\nu}} \sigma_{i_\nu}^{\alpha}h^{eff}_{ijk}\right]
 \nonumber \\ && -\frac{\beta J^2}{2}\left[\sum_{\alpha}{q}_{\alpha\alpha}(\sigma_{\nu}^{\alpha})^2
 +\sum_{\alpha\gamma}\frac{q_{\alpha\gamma}}{2}\sigma_{\nu}^{\alpha}\sigma_{\nu}^{\gamma}\right]
\label{Heff}\end{eqnarray}
where $h^{eff}_{ijk}= J_0(m^{\sigma_{i} \sigma_{j}}+m^{\sigma_{i}\sigma_{k}})$ with $i$ and $j$ (or $k$) referring to pairs of spins at the cluster boundary $\bar{\nu}$ that interact with spins at the same neighbor cluster. 
 The effective field can also be expressed as in Eq. (\ref{effpure}). This approach is applicable in both cluster shapes depicted in Fig. (\ref{CCMF}).   $m^{\sigma_{i} \sigma_{j}}$ represents mean fields that depend on the spin states of sites $i$ and $j$ of cluster $\nu$.

Finally,  Eq. \ref{Heff} represents a single cluster inter-replica coupling problem.
The simplest approach to this coupling is the replica-symmetric (RS) solution that considers $q_{\alpha\gamma}=q$ and $q_{\alpha\alpha}=\bar{q}$. 
The resulting  effective Hamiltonian of the cluster $\nu$ becomes
\begin{eqnarray}
 H^{RS}_{eff} (x) &=& - J_0 \sum_{(i,j)}^{n_s}\sigma_{i_\nu} \sigma_{j_\nu} -   \sum_{i\in\bar{\nu}}\sigma_{i_\nu}h_{ijk}^{eff}
  \nonumber \\  && -\frac{\beta J^2}{2} (\bar{q}-q) \sigma_{\nu}^2 - J\sqrt{q} x \sigma_{\nu},
\label{HeffRS}
\end{eqnarray}
with
\begin{equation}
q=\int Dx \left( \frac{\mbox{Tr } \sigma_\nu \exp{(-\beta H^{RS}_{eff}}) }{\mbox{Tr}  \exp{(-\beta H^{RS}_{eff}} )}\right)^2,
\label{q}\end{equation}
\begin{equation}
\bar{q}=\int Dx \frac{\mbox{Tr } \sigma_\nu\sigma_\nu \exp{(-\beta H^{RS}_{eff}}) }{\mbox{Tr}  \exp{(-\beta H^{RS}_{eff}} )}
\label{qb}\end{equation}
and
\begin{equation}
m^{ss'}=\int Dx \frac{\mbox{Tr } \sigma_{i_\nu'} \exp{(-\beta H^{RS}_{\nu'}(s,s')}) }{\mbox{Tr}  \exp{(-\beta H^{RS}_{\nu'}(s,s')} )},
\label{mss}\end{equation}
 where $Dx=e^{-x^2/2}/\sqrt{2\pi}$ and $H^{RS}_{\nu'}(s,s')$ is the effective Hamiltonian of the neighbor cluster $\nu'$ (defined in Eq. (\ref{Heff_neig})) for a given spin configuration $s$ and $s'$ from connected sites of the neighbor cluster $\nu$. It means that computing $q$, $\bar{q}$ and $m^{s s'}$ (with $s= \pm 1$ and $s'=\pm 1$) in a self-consistent way, we get the effective model given by Eq. (\ref{HeffRS}).
Therefore, we solve numerically the system composed by Eqs. (\ref{HeffRS}), (\ref{q}), (\ref{qb}) ,(\ref{mss}) and (\ref{Heff_neig}) to obtain the single cluster model (\ref{HeffRS}). Then, Eq.
(\ref{HeffRS}) is used to evaluate all other thermodynamic quantities as internal energy ($u$), specific heat ($C_m$), entropy ($S$), magnetization and  susceptibility $\chi$ (see \ref{termo}). However, the RS solution can be unstable within the  CSG phase as it is found by the de Almeida-Thouless analysis  given by $\lambda_{AT}$ in Eq. (\ref{lat}). 

\section{Results and Discussion}\label{res}
\begin{figure}[t]
\center{\includegraphics[width=0.99\columnwidth]{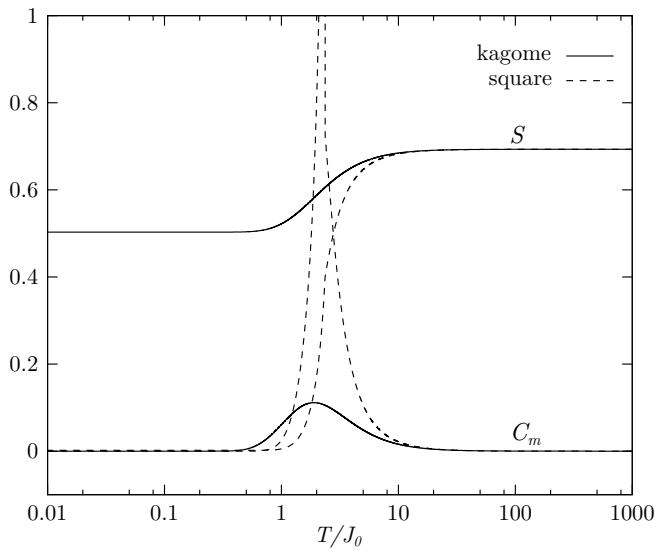}}
\caption{Entropy ($S$) and specific heat ($C_m$) as a function of temperature for square and kagome lattices in the clean limit with $J_0=-1$.}  
\label{clean_s_Cm}  \end{figure}

In the absence of disorder ($J=0$), our approach falls into the 2D Ising model, with the numerical results obtained by solving the self-consistent equations of the CCMF method (Eqs. (\ref{Heff_neig1}) and (\ref{heffpure})). Then we get the effective model (\ref{effpure}) to derive the thermodynamic quantities.
For  AF couplings ($J_0<0$), we consider a sub-lattice structure for 
the square system, that presents staggered magnetization $m_s$ characterizing the AF long-range order.  We found the N\'eel temperature $T_N/J_0=2.362$ 
\cite{prezimmer2014}, which is very close to the exact one ($ T_N/J_0=2.269$)\cite{PhysRev.65.117}. 
As a consequence of the AF order, the entropy goes to zero when $T \to 0$ and the specific heat $C_m$ shows a discontinuity at $T_N$ (see dashed lines of Fig. \ref{clean_s_Cm}).

 \begin{figure*}[t]
\center{\includegraphics[angle=0,width=0.99\textwidth]{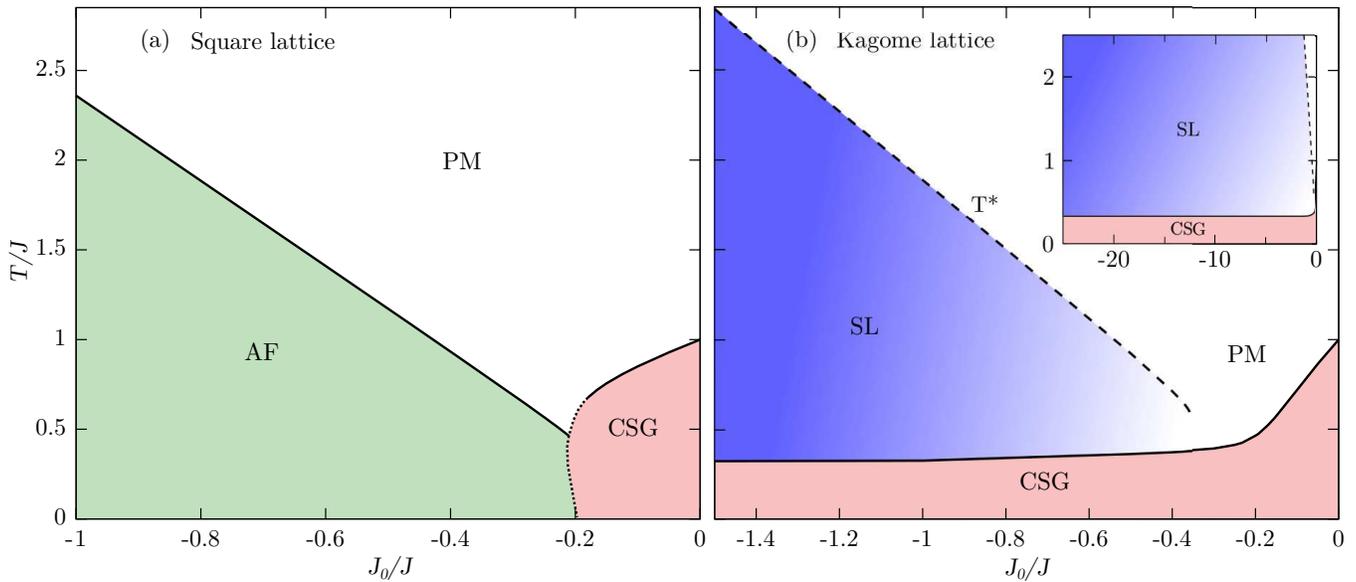}}
\caption{(color online) Phase diagrams of $T/J$ vs. $J_0/J$ for the (a) square and (b) kagome lattices. Solid lines indicate continuous phase transitions. In (a) the dotted line is the CSG phase stability and the dashed line in (b) indicates the crossover between PM and SL behavior. Inset shows the behavior of $T_f$ for $J<<|J_0|$. For convenience we set the disorder strength $J$ such that $T_f/J=1$ for $J_0=0$. }  
\label{phase_diagrams_square}  \end{figure*}

However, the magnetic behavior of the kagome lattice is completely different.
The strong GF  avoids a conventional AF state, and the system remains disordered even at $T=0$. In this case, a classical SL state with macroscopic degeneracy is observed at low temperatures. For instance, Fig. \ref{clean_s_Cm} exhibits important signatures of this classical SL regime, i.e., the entropy plateau and the low $C_m$ at low temperatures \cite{Rep_Prog_Balents_2016}. The specific heat shows a maximum at $T/J_0 \approx 2$ which could be used as an estimation for the crossover temperature $T^*$ between the high temperature paramagnetic (PM) state and the low-temperature cooperative PM one, i. e., the SL state \cite{PhysRevLett.106.207202,PhysRevB.93.214410_Canals_2016,canals_natcom,PhysRevB.94.014429}.
Furthermore, the CCMF method leads to a very accurate result for the ground-state entropy ($S_{res}=0.503$) \cite{Schmidt2015416}, when compared to the exact one  for the kagome system ($S_{res}^{exact}=0.502$) \cite{Kano01081953}.

\subsection{The disordered square  lattice}\label{disorder}

 \begin{figure}
\center{\includegraphics[angle=0,width=0.99\columnwidth]{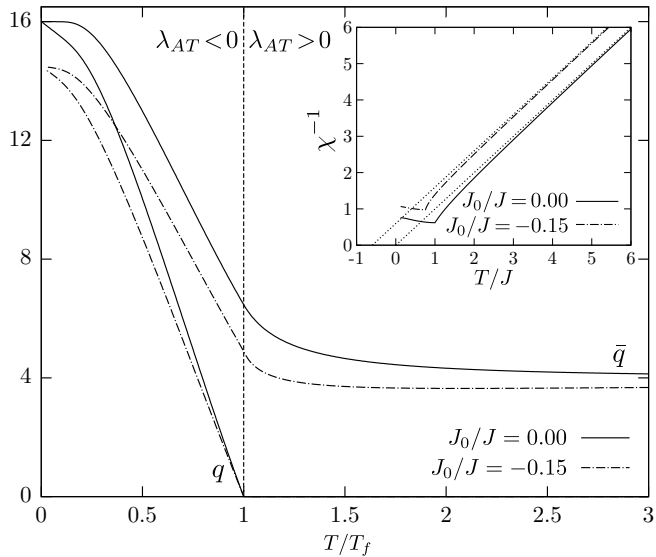}}
\caption{Order parameters $q$ and $\bar{q}$ as a function of normalized temperature ($T/T_f$) for  two strength of short-range interactions in the square lattice. Inset 
exhibits the inverse of magnetic susceptibility $\chi^{-1}$ with its linear extrapolation from higher temperatures (dotted line). }  
\label{op_sg}  \end{figure}

The presence of disordered interactions can introduce the CSG phase ($q>0$ with $\lambda_{AT}<0$). For instance, Fig. \ref{phase_diagrams_square} (a) exhibits a phase diagram for the disordered square system, in which a phase transition from the PM to the CSG behavior is found at the freezing temperature $T_f$.  In particular, the replica-symmetric solution is unstable ($\lambda_{AT}<0$) in the whole CSG phase.
The AF interactions depress the freezing temperature until a sufficiently large intensity of $J_0/J$, in which the AF order ($m_s>0$ with $q=0$ and $\lambda_{AT}>0$) becomes stable.
A discontinuous phase transition between the cluster SG and the AF is observed and the stability limit of the CSG phase is indicated by the dotted line in Fig. \ref{phase_diagrams_square} (a). These results indicate a strong competition between antiferromagnetism and CSG in the square lattice. This competition is introduced by AF interactions that bring a cluster compensation mechanism, reducing the cluster magnetic moment, which is against the CSG stabilization.

Figure \ref{op_sg} helps to discuss this interplay between the short-range interaction and the  CSG state, in which the behavior of $\bar{q}$ becomes important. 
$\bar{q}$ represents the replica diagonal elements 
and can be interpreted as the average of the cluster magnetic moment. 
Different from the canonical SK model ($\bar{q}=1$), here  $\bar{q}$ depends on $T/J$ and $J_0/J$ and it affects the thermodynamics of the PM phase. 
For instance, AF interactions reduce $\bar{q}$ (see Fig. \ref{op_sg}), introducing a competitive scenario that leads to the reduction of $T_f$, as shown in Fig. \ref{phase_diagrams_square} (a).

The signature of short-range couplings can also be observed from the magnetic susceptibility behavior. 
For instance, the Curie-Weiss temperature $\theta_{CW}$ is evaluated 
from $\chi^{-1}$, which follows the Curie-Weiss law at higher temperatures (see the inset of Fig. \ref{op_sg}). 
The negative 
$\theta_{CW}$ found for $J_0/J=-0.15$ 
indicates an antiferromagnetic 
bias. 
The $\theta_{CW}$ could also be used to evaluate a parameter $f=|\theta_{CW}|/T_c$ related to GF, where $T_c$ refers to the transition temperature to any ordered state \cite{Ramirez_GF_1994}. 
A strong suppression of ordering due to GF is indicated in general by $f \gtrsim 5$ \cite{Ramirez_GF_1994, Balents_SL}. For the disordered square lattice, we found $f\leqslant 1$, which is a consequence of the absence of GF.

\subsection{The disordered kagome lattice}\label{kagomedisorder}

 \begin{figure}[ht]
\center{\includegraphics[angle=0,width=0.99\columnwidth]{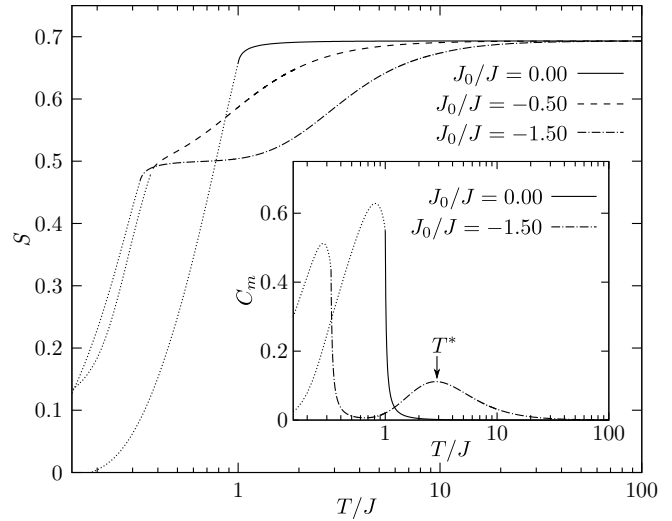}}
\caption{ Entropy versus temperature for several values of short-range couplings on the kagome lattice.  Inset shows the specific heat as a function of temperature. Dotted lines indicate RS unstable regions ($\lambda_{AT}<0$).}  
\label{fig2}  \end{figure}

In the following, we discuss the disordered AF kagome results. The phase diagram presents a particularly interesting scenario due to GF, differing it from the square lattice  (see Fig. \ref{phase_diagrams_square} (b)).
For a weak AF coupling the $T_f/J$ is reduced as the intensity of $J_0/J$ increases. At intermediary AF coupling ($J_0/J \approx -0.5$) the $T_f$ becomes weakly dependent on the intensity of $J_0$. 
For higher values of $J_0$, the $T_f$ becomes uniquely dependent on the strength of $J$, as it is exhibited in the inset of Fig. \ref{phase_diagrams_square} (b). It means that the CSG is always the ground state if any intercluster disorder is present. 
 In particular, this result agrees qualitatively with the analytical and numerical findings for the AF pyrochlore lattice in a weak disorder regime, in which $T_f$ is proportional to the amplitude of the interaction strength deviation when both gaussian and homogeneous disorder distributions are considered \cite{PhysRevB.81.014406}. In this sense, we believe that the $T_f$ dependence on $J$ can be robust for other types of disorder in the cluster kagome model under study. However, further investigations are needed to account for this point.
As we will discuss below, for a large enough intensity of AF coupling, we find signatures of classical SL onset above $T_f$, which is indicated by the dashed line ($T^*$) in the phase diagram \ref{phase_diagrams_square}(b). 

A detailed thermodynamic analysis helps to characterize this interplay between GF and disorder. For instance, Fig. \ref{fig2} shows the entropy for different intensities of  $J_0/J$. For weak AF couplings, the entropy exhibits a usual high-temperature plateau and a drop close to $T_f$. On the other hand, for $J_0/J\lesssim -0.5$ the entropy shows a second plateau at intermediary temperatures. This plateau of $S \approx 0.5$ occurs between $T_f$ and a second specific heat maximum (see the inset of Fig. \ref{fig2}). A low specific heat is also observed in the same range of temperature of the entropy plateau, resembling the results for the clean  AF kagome.
We identify the second maximum in $C_m$ as 
the crossover temperature ($T^*$) between the high-temperature  PM state and the classical  SL regime. Moreover, $T^*$ becomes linearly dependent of $J_0$ for $J_0/J\lesssim -0.5$ reinforcing that the second $C_m$ maximum can be related to the onset of the SL behavior. Therefore, specific heat and entropy results indicate that the region of intermediary temperature $T_f<T<T^*$ is ruled by  GF when $J_0/J\lesssim -0.5$.

 \begin{figure}
\center{\includegraphics[angle=0,width=0.99\columnwidth]{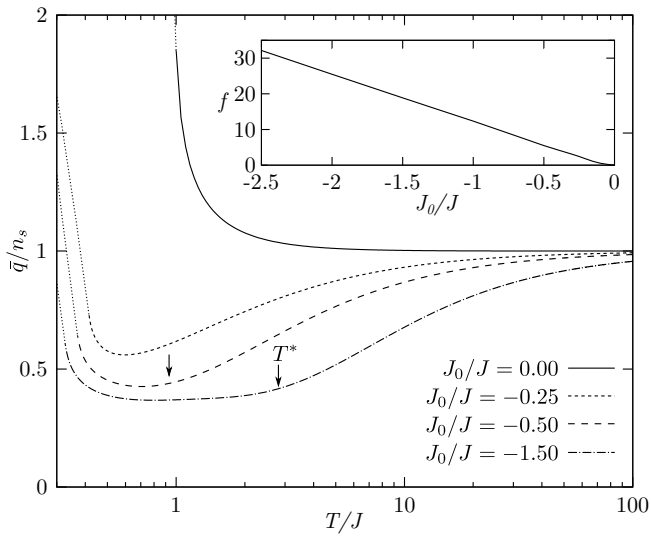}}
\caption{Temperature dependence of the normalized cluster magnetic moment $\bar{q}/n_s$ for various AF couplins in the kagome lattice. Dotted lines indicate RS unstable regions ($\lambda_{AT}<0$). Inset shows the frustration parameter as a function of $J_0/J$.}  
\label{fig4222}  \end{figure}

For weak AF couplings, $\bar{q}$ behaves in a similar way to that observed in the square lattice (see Fig. \ref{fig4222} for $J_0/J=-0.25$). 
However, $\bar{q}$ becomes weakly dependent on the temperature within the SL regime (see  Fig. \ref{fig4222} for $J_0/J=-0.50$ and $-1.50$). 
In addition, 
$\bar{q}$ is minimum in this region indicating that the AF couplings overcome the temperature effects. 
It means that the system is strongly affected by the short-range  AF couplings, which leads the cluster magnetic moment to a minimum value. 
However, some of the many degenerated states introduced by GF lead to uncompensated cluster moment, reflecting in a finite $\bar{q}$.
Therefore, this $\bar{q}$ result for $J_0/J\lesssim -0.5$ can be understood as GF effect, which is consistent with the behavior of the frustration parameter in the inset of Fig. \ref{fig4222}. 
For instance, for $J_0/J \approx -0.5$ the frustration parameter reaches the value considered as a signature of GF ($f>5$) \cite{Ramirez_GF_1994}.
It is important to remark that a weak disorder leads the system to choose uncompensated cluster states from the multi-degenerate scenario introduced by GF. 
Despite small, the cluster magnetic moment is enough to activate disorder effects. This is the mechanism that favors the  CSG state to appear at lower temperatures. 

\begin{figure}[ht]
\center{\includegraphics[angle=0,width=0.99\columnwidth]{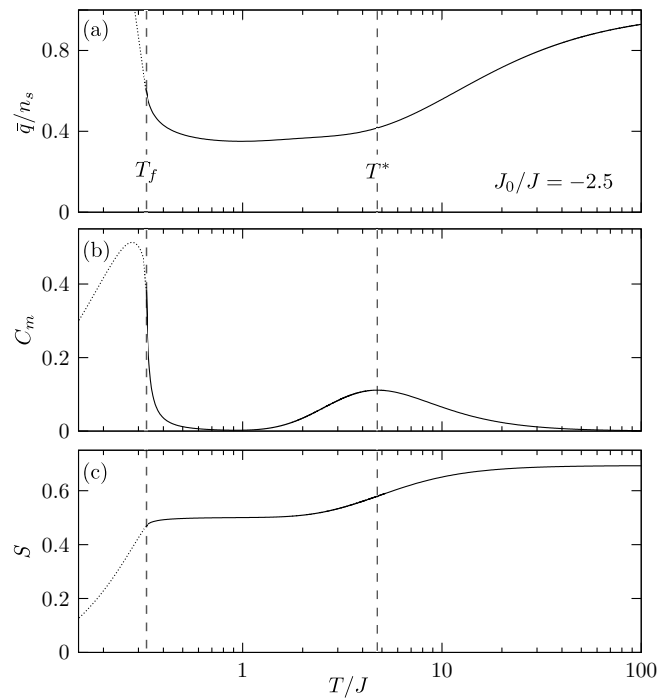}}
\caption{Temperature dependence of the (a) normalized cluster magnetic moment $\bar{q}/n_s$, (b) magnetic specific heat $C_m$, and (c) entropy $S$ for $J_0/J=-2.5$ in the kagome lattice. Dotted lines indicate RS unstable regions ($\lambda_{AT}<0$).  }
\label{res_250}  \end{figure}

In order to enforce our physical picture for a regime of strong GF ($f \approx 32$ as shown in the inset of Fig. \ref{fig4222}), we present the temperature dependence of $\bar{q}$, $C_m$ and $S$ in Fig. \ref{res_250} for $J_0/J = -2.5$. 
When temperature is reduced from the high-temperature regime, $\bar{q}$ exhibits a drop, which coincides with an increasing specific heat and an entropy release.
Below the maximum of $C_m$ ($T^*$), $\bar{q}$ becomes weakly dependent on the temperature in the same region where the finite entropy plateau occurs. 
However, when temperature is reduced below $T/J \approx 1$, an increase in the magnetic specific heat and a second entropy drop can be observed. 
This results could be understood as an effect of the intercluster disorder, that breaks the degeneracy favoring the CSG state at very low temperatures. 
In this scenario, $\bar{q}$ increases when temperature is reduced even before the replica symmetry broken takes place. 
It is important to remark that in this weak disorder regime $T_f$ is ruled by $J$ and not by $J_0$.

\section{Conclusion}\label{conc}
We study the effects of disorder in the Ising kagome lattice. We assume that disorder introduces a relevant degree of freedom associated with the presence of clusters.
We found that a regime of strong geometrical frustration brings classical spin-liquid (SL) signatures at the same time that an infinitesimal disorder leads to a cluster spin-glass (CSG) ground state. 
The strong antiferromagnetic  (AF) couplings introduce a high degeneracy reflecting in an entropy plateau at finite temperature followed by a specific heat maximum related to the SL regime. 
However, this frustration leads to uncompensated clusters potentializing the disorder effects. This is the mechanism that allows a low-temperature CSG state driven by any amount of disorder. In this scenario, the complex ergodicity breaking can be preceded by the SL behavior as temperature decreases.
For comparison, we also study the disordered AF square lattice, in which geometrical frustration is absent. In contrast, a finite value of the  AF couplings eliminates the  CSG phase, giving rise to an  AF state (see Fig. \ref{phase_diagrams_square}). 
It corroborates with the picture that both clusters and geometrical frustration are the driven forces of  the glassy behavior found in systems with very low levels of disorder.

Although we consider a particular cluster mean-field model,  we identify a physical mechanism that could be present in several real systems, particularly for the compound Co$_3$Mg(OH)$_6$Cl$_2$,  in which signatures of a collective paramagnetic state are observed at temperatures above the glassy behavior \cite{zheng2012}. 
In this compound,
an entropy plateau of $S\approx 0.5$ followed by a large entropy drop close to $T_f$  was reported. Moreover, our findings suggest that very small clusters could be present, 
as supported by neutron diffraction and muon spin rotation/relaxation results \cite{zheng2012}, providing hypersensitivity to  the freezing process. 
As a consequence, glassiness is expected in this material even at extremely low levels of disorder. 
However, further experimental investigations are still needed to account for the nature of the glassy state and SL behavior indeed. 
 Moreover, it will be welcome additional analytical and numerical studies to account for different types of bond disorder and quantum fluctuations in the present cluster system.

\begin{acknowledgments}
This work was partially supported by the brazilian 
agency Conselho Nacional de Desenvolvimento Cient\'ifico e Tecnol\'ogico (CNPq). 
We acknowledge E. V. Sampathkumaran and C. Lacroix for useful discussions.  
\end{acknowledgments}

\appendix \section{CCMF decoupling}\label{CCMF_dec}
We consider the short-range couplings 
given by $H_{J_0}^{\alpha}$ (see Eq. \ref{eq2}):
\begin{equation}
 H_{J_0}^{\alpha}= -\sum_{\nu}^{N_{cl}}\sum_{(i,j)}^{n_s}J_0 \sigma_{i_\nu}^{\alpha} \sigma_{j_\nu}^{\alpha}   - \sum_{(i_\nu, j_\lambda)} J_{0}\sigma_{i_\nu}^{\alpha} \sigma_{j_\lambda}^{\alpha},
 \label{appA1}\end{equation}
which refers to interactions in the same replica. In this way, we suppress the replica index and treat the intercluster interaction (second term) within the CCMF approach: 
 \begin{equation}
 \sum_{(i_\nu, j_\lambda)} \sigma_{i_\nu}^{\alpha} \sigma_{j_\lambda}^{\alpha}\approx
 \sum_{i_\nu\in \bar{\nu}}\sigma_{i_\nu}(m^{\sigma_{i_\nu}\sigma_{j_\nu}}_{ij}+m^{\sigma_{i_\nu}\sigma_{k_\nu}}_{ik}),
\end{equation}
where $i_\nu$ and $j_\nu$ (or $k_\nu$) are sites of the cluster boundary ($\bar{\nu}$) and they are neighbors of the same clusters. 
For instance, for the square lattice (see Fig. \ref{CCMF} (a)) $i_\nu=1$, $j_\nu=2$ and $k_\nu=3$, and for the kagome lattice (see Fig. \ref{CCMF} (b)) $i_\nu=1$, $j_\nu=5$ and $k_\nu=2$. 
In addition, $m^{\sigma_{i_\nu}\sigma_{j_\nu}}_{ij}$ represents the mean fields $m^{++}_{ij},$ $m^{+-}_{ij},$ $m^{-+}_{ij}$ and $m^{--}_{ij}$, which are associated to the four possible spin configurations of the sites $i_\nu$ and  $j_\nu$: $|\uparrow\uparrow\rangle$, $|\uparrow\downarrow\rangle$, $|\downarrow\uparrow\rangle$ and $|\downarrow\downarrow\rangle$, respectively. 
In general, two set of these mean fields should be evaluated for each site at the cluster boundary ($m^{\sigma_{i_\nu}\sigma_{j_\nu}}_{ij}$ and $m^{\sigma_{i_\nu}\sigma_{k_\nu}}_{ik}$), without explore the symmetries. For instance, for the kagome system exhibited in Fig. \ref{CCMF} (b), we should evaluate 48 mean fields. However, the topological equivalence of the boundary sites can be used, reducing the numerical cost of the method to find only four mean fields: $m^{++},$ $m^{+-},$ $m^{-+}$ and $m^{--}$. In order to check it, we evaluated all the 48 mean fields, obtaining the same results.

Therefore, in the CCMF framework, we can express 
\begin{eqnarray}
 \sigma_{i_\nu} m^{\sigma_{i_\nu}\sigma_{j_\nu}} &=& \sigma_{i_\nu}[(1+\sigma_{i_\nu}) (1+\sigma_{j_\nu})m^{++}
 \nonumber \\ && +(1+\sigma_{i_\nu})(1-\sigma_{j_\nu}) m^{+-}
 \nonumber \\ && +(1-\sigma_{i_\nu})(1+\sigma_{j_\nu})m^{-+}
 \nonumber \\ && + (1-\sigma_{i_\nu})(1-\sigma_{j_\nu})m^{--}]/4.
\end{eqnarray}
Thus,
\begin{eqnarray}
 \sigma_{i_\nu}(m^{\sigma_{i_\nu}\sigma_{j_\nu}}+m^{\sigma_{i_\nu}\sigma_{k_\nu}}) &=& [(\sigma_{i_\nu}\sigma_{j_\nu}+\sigma_{i_\nu}\sigma_{k_\nu})C  + 2D
  \nonumber \\ &&  + (\sigma_{j_\nu}  +\sigma_{k_\nu})B \nonumber \\ &&  + 2\sigma_{i_\nu} A  ]/4, \label{A4}
\end{eqnarray}
where $A=m^{++}+m^{+-}+m^{-+}+m^{--}$, $B=m^{++}-m^{+-}-m^{-+}+m^{--}$, $C=m^{++}-m^{+-}+m^{-+}-m^{--}$ and $D=m^{++}+m^{+-}-m^{-+}-m^{--}$.
In this way, using Eq. (\ref{A4}), it is possible to rewrite Eq. (\ref{appA1})  as
\begin{equation}
H_{J_0}= -J_0 \sum_{(i,j)}^{n_s}\sigma_{i_{\nu}} \sigma_{j_{\nu}} 
-\sum_{i \in \bar{\nu}}\sigma_{i_{\nu}}h^{eff}_{ijk}
\label{heffJ0}\end{equation}
where 
\begin{equation}
h^{eff}_{ijk}=J_0[m^{++}+m^{--}+(\sigma_{j_\nu}+\sigma_{k_\nu})C/4]
\label{effpure}\end{equation}
depends on the mean fields and the spin states assumed by the boundary sites $j_\nu$ and $k_\nu$ that are neighbors of $i_\nu$. In particular, the spin states dependence of the effective fields is one of the differences between the CCMF approach and the standard cluster mean-field method\cite{Yamamoto}. This dependence occurs as a consequence of the effective renormalization of the intercluster interactions introduced by the mean fields, which provide corrections to the standard mean-field treatment. 

In order to obtain the set of mean fields, we consider the nearby connected cluster $\nu'$ (see Fig. \ref{CCMF} (b)). For the kagome lattice, the spins of sites $2'$ and $8'$ interact with each one of the possible spin configurations of sites 5 and 11 of cluster $\nu$, while the other intercluster interactions are replaced by  mean fields. The mean field $m^{s s'}$ is the average value of the $\sigma_{2_{\nu'}}$ when the spin configurations of sites 5 and 11 is $|s s'\rangle$:
\begin{equation}
 m^{s s'}= \langle \langle \sigma_{2'}\rangle \rangle_{H_{\nu'}^{eff}(s,s')}
\end{equation}
where $H_{\nu'}^{eff}(s,s')= H_{dis} + H_{J_0 }^{\nu'}(s,s')$
\begin{eqnarray}
H_{J_0 }^{\nu'}(s,s')&=&  -J_0 \sum_{(i,j)}^{n_s}\sigma_{i_{\nu'}} \sigma_{j_{\nu'}} 
%
-\sum_{ { i \in \bar{\nu'}}; \left(i\neq 2',8'\right)  }
\sigma_{i_{\nu'}}h^{eff}_{ijk} 
\nonumber \\ && - \sigma_{2'}h_{2'1'}-\sigma_{8'}h_{8'12'}  \nonumber \\ && -J_0[s\sigma_{2'}+s'\sigma_{8'}],
\label{Heff_neig1}\end{eqnarray}
with $h_{2'1'}=J_0[(m^{++}+m^{--})/2 +\sigma_{1'}C/4]$,
$H_{\nu'}^{eff}$ is the effective Hamiltonian of the cluster $\nu'$ and $H_{dis}=-\frac{\beta J^2}{2}[\sum_{\alpha}{q}_{\alpha}(\sigma_{\nu'}^{\alpha})^2
 +\sum_{\alpha\gamma}\frac{q_{\alpha\gamma}}{2}\sigma_{\nu'}^{\alpha}\sigma_{\nu'}^{\gamma}]$. Within the replica symmetric solution, $H_{\nu'}^{eff}$ becomes
\begin{eqnarray}
H_{\nu'}^{RS}(s,s')&=&-\frac{\beta J^2}{2} (\bar{q}-q) \sigma_{\nu'}^2 - J\sqrt{q} x\sigma_{\nu'} \nonumber \\ && + H_{J_0 }^{\nu'}(s,s').
\label{Heff_neig}\end{eqnarray}

In particular, the effective single cluster model for the free-disorder limit is given by ($J=0$) Eq. \ref{heffJ0} with  
\begin{equation}
m^{ss'}= \frac{\mbox {Tr} \sigma_{2'} \mbox{e}^{-\beta H_{J_0}^{\nu'}(s,s')}}{\mbox{Tr e}^{-\beta H_{J_0}^{\nu'}(s,s')}}
\label{heffpure}\end{equation}
where $H_{J_0}^{\nu'}(s,s')$ is defined in Eq. (\ref{Heff_neig1}).

The square lattice is also considered within the CCMF procedure. As in the kagome case, it presents two spin interactions between nearby clusters (see Fig. \ref{CCMF} (a)). However, this system can exhibit an antiferromagnetic state characterized by the staggered magnetization in a two sublattice structure, as depicted in Fig \ref{CCMF} (a).
Therefore, it is necessary to compute a mean-field set for each sublattice.
For instance, the mean fields that act in the sublattice composed by the sites $2$ and $3$ (or $1$ and $4$) are computed from the average value of $\sigma_{1_\nu'}$ (or $\sigma_{3_\nu'}$), by using parameters $s$ and $s'$, as indicated in  Fig \ref{CCMF} (a). Details of the CCMF approach for N\'eel antiferromagnets are provided in Ref. \cite{PhysRevE.89.062117}.

 \section{Thermodynamic quantities and RS solution analysis}\label{termo}

The internal energy per cluster $U$ can be computed from Eq. (\ref{freeenergy})
\begin{eqnarray}
U &=& \lim_{n\rightarrow 0}\left[\frac{1}{N_{cl}n}\frac{\mbox{Tr}(\sum_{\alpha}H^{\alpha}_{J_0})e^{-\beta H_{eff}}}{\mbox{Tr}e^{-\beta H_{eff}}}\right.
\nonumber \\ && \left. -\frac{\beta J^2}{2n}\left(\sum_{\alpha}(q^{\alpha\alpha})^2 +\sum_{\alpha\gamma}(q^{\alpha\gamma})^2 \right) \right],
\end{eqnarray}
where $H_{eff}$ is defined in Eq. (\ref{Hav}). Using replica-symmetric solution and the CCMF approach, we obtain the following internal energy per site $u=U/n_s$
\begin{eqnarray}
u &=&  -\frac{\kappa J_0}{n_s} \int Dx
\left\langle \sum_{(i,j)}^{n_s}\sigma_{i_\nu}\sigma_{j_\nu} \right\rangle_{H_{eff}^{RS}(x)} \nonumber \\ && + \frac{\beta J^2}{2n_s }(q^2-\bar{q}^2), 
\end{eqnarray}
where the average is computed with $H_{eff}^{RS}(x)$ defined in Eq. (\ref{HeffRS}). $\kappa$ accounts for the intercluster couplings evaluated with the CCMF, in which $\kappa=4/3$ (or $\kappa=2$) for the kagome (or the square) lattice.  The specific heat per site can be derived as $c_v=\frac{d u}{dT}$ and the entropy per site is given by integration of $c_v/T$:
\begin{equation}
S = \int_{0}^{T} \frac{c_v}{T'} dT'= \ln (2)- \int_{T}^{\infty} \frac{c_v}{T'} dT'.
\end{equation}

The magnetization can be computed from 
\begin{equation}
m = \frac{1}{n_s}\int Dx \frac{\mbox{Tr} \sum_{(i)}\sigma_i e^{-\beta H_{eff}^{RS}(x)}}{\mbox{Tr}e^{-\beta H_{eff}^{RS}(x)}}.
\end{equation}
For the square lattice, we can also calculate the staggered magnetization given by $m_s=(m_a-m_b)/2$, where $m_a$ ($m_b$) corresponds to the magnetization of the sublattice $a$ ($b$). The magnetic susceptibility is obtained by numerical derivation of the magnetization per site $\chi = \left(\frac{ \partial m }{ \partial h}\right)_{h=0}$.

The stability analysis of the replica-symmetric solution can be performed by de Almeida-Thouless eigenvalues \cite{AT}. In particular, the replicon mode assumes the expression
\begin{eqnarray}
\lambda_{AT}&=& 
 -(\beta J)^4\int Dx \left(\langle \sigma_\nu\sigma_\nu \rangle_{H_{eff}^{RS}(x)}-\langle \sigma_\nu \rangle^2_{H_{eff}^{RS}(x)} \right)^2 
 \nonumber \\ && + (\beta J)^2 .
\label{lat}\end{eqnarray}
It is important to remark that, different from the canonical Ising spins ($\langle \sigma_i\sigma_i \rangle=1$), here $\langle \sigma_\nu\sigma_\nu \rangle_{H_{eff}^{RS}(x)}$ have to be considered.

\section*{References}\label{ref}

\bibliography{References}{}
\bibliographystyle{iopart-num}

\end{document}